\documentclass[aps,prl,showpacs,twocolumn,preprintnumbers,superscriptaddress,nofootinbib,floatfix,10pt]{revtex4-1}
\usepackage{amsfonts,amssymb,stmaryrd,latexsym,amsmath,braket}
\usepackage{graphicx,subfigure}
\usepackage{comment}
\usepackage{times}
\usepackage{slashed}
\usepackage{bm}
\usepackage{braket}
\usepackage{chapterbib}
\usepackage{color}

\makeatletter
\let\saved@includegraphics\includegraphics
\AtBeginDocument{\let\includegraphics\saved@includegraphics}
\makeatother

\graphicspath{{figures/}}
\begin{document}

\title{{{\em Ab initio} calculation of the alpha-particle monopole transition form factor}}


\author{Ulf-G.~Mei{\ss}ner}
\affiliation{Helmholtz-Institut f\"{u}r Strahlen- und Kernphysik and Bethe Center for Theoretical Physics, Universit\"{a}t Bonn, D-53115 Bonn, Germany}
\affiliation{Institut f\"{u}r Kernphysik, Institute for Advanced Simulation and J\"{u}lich Center for Hadron Physics, Forschungszentrum J\"{u}lich, D-52425 J\"{u}lich, Germany}
\affiliation{Tbilisi State University, 0186 Tbilisi, Georgia}

\author{Shihang Shen}
\affiliation{Institute for Advanced Simulation and Institut f\"{u}r Kernphysik, Forschungszentrum J\"{u}lich, D-52425 J\"{u}lich, Germany}

\author{Serdar Elhatisari}
\affiliation{Faculty of Natural Sciences and Engineering, Gaziantep Islam Science and Technology University, Gaziantep 27010, Turkey}
\affiliation{Helmholtz-Institut f\"{u}r Strahlen- und Kernphysik and Bethe Center for Theoretical Physics, Universit\"{a}t Bonn, D-53115 Bonn, Germany}

\author{Dean Lee}
\affiliation{Facility for Rare Isotope Beams and Department of Physics and Astronomy, Michigan State University, East Lansing, MI 48824, USA}

\begin{abstract}
We present a parameter-free {\em ab initio} calculation of the $\alpha$-particle monopole transition form factor
in the framework of nuclear lattice effective field theory. We use a  minimal nuclear interaction that
was previously used to reproduce the ground state properties of light nuclei, medium-mass nuclei, and neutron matter simultaneously
with no more than a few percent error in the energies and charge radii.  The  results for the monopole transition form factor are in good agreement with recent precision data from Mainz. 
\end{abstract}
\maketitle
\date{today}

\section{Introduction}

The $^4$He nucleus, the $\alpha$-particle, is considered to be a benchmark nucleus for our understanding of  the
nuclear forces and the few-body methods to solve the nuclear $A$-body problem~\cite{Kamada:2001tv}.
The attractive nucleon-nucleon interaction makes this highly symmetric four-nucleon system
enormously stable. Furthermore, its first excited state has the same quantum numbers as the 
ground state, $J^P=0^+$ with $J(P)$ the spin (parity), but is located about 20~MeV above the ground state.
This large energy of the first quantum excitation makes the system difficult to perturb. 
This isoscalar monopole resonance of the $^4$He nucleus presents a challenge to
our understanding of nuclear few-body systems and the underlying nuclear forces \cite{Bacca:2012xv}.
Within pionless effective field theory, the ground state and the first excited state of the alpha particle
could already be reproduced in Ref.~\cite{Kirscher:2018dwo}, although with some uncertainty in the position
of the first excited state to the proton-triton threshold.
The recent precision measurement of the corresponding transition form factor of the first excited state to the
ground state at the Mainz Microtrom MAMI~\cite{Kegel:2021jrh} compared with  {\em ab initio}
calculations based on the Lorentz-integral transformation method~\cite{Efros:1994iq}
using phenomenological potentials as well as potentials based on chiral effective field theory, e.g.~\cite{Beane:2000fx,Epelbaum:2008ga,Machleidt:2011zz,Hammer:2019poc,Epelbaum:2019kcf},
revealed sizeable discrepancies as shown in Fig.~3 of Ref.~\cite{Kegel:2021jrh}. 

These new results have spurred a number of theoretical investigations, stressing
especially the role of the continuum when including the resonant state which is located close to
the two-body breakup threshold~\cite{EE,Michel:2023ley}. In particular, Ref.~\cite{Michel:2023ley}
showed that employing an explicit coupled-channel representation of the no-core Gamow shell model
with the $^3$H+$p$, $^3$He$+n$ and $^2$H$+^2$H reaction channels allows to reproduce the Mainz data.
In that paper, the effects of three-nucleon forces were neglected.  We remark that the Mainz data
are also reproduced by the pioneering work of Ref.~\cite{Hiyama:2004nf}, which also pointed out the
importance of the loosely bound $3N$+$N$ system, where $N$ denotes a nucleon. However, as noted
in Ref.~\cite{Kegel:2021jrh}, that calculation does not reproduce the low-energy data, more precisely, the two first
parameters in the low-momentum expansion of the transition form factor.
It was also pointed out that  the shape of the transition density obtained from the data in Ref.~\cite{Kegel:2021jrh}
is significantly
different from that obtained theoretically in the literature~\cite{Kamimura:2023tvl}.

Here, we will use the framework of nuclear lattice effective field theory (NLEFT) to present an {\em ab initio} solution
to the problem~\cite{Lee:2008fa,Lahde:2019npb}. In particular, we want to address the issue whether one possibly misses parts of the nuclear force
which,
given a simple spin-0 and isospin-0 nucleus like $^4$He, would be rather striking. We make use of a so-called minimal nuclear interaction, that has been successfully used
to describe the gross properties of light and medium-mass nuclei and the equation of state of
neutron matter to a few percent accuracy~\cite{Lu:2018bat}. It was used in  nuclear
thermodynamics calculations~\cite{Lu:2019nbg} and {\em ab initio} studies of clusters in hot dilute matter using the
method of light-cluster distillation~\cite{Ren:2023ued}. 
A similar action was also successfully applied to
investigate  the emergent geometry and intrinsic cluster structure of the low-lying states of $^{12}$C~\cite{Shen:2022bak}. 
In particular, the transition form factor from the Hoyle state to the ground state measured at Darmstadt~\cite{Chernykh:2007zz}
could be excellently reproduced without any parameter adjustment.
In this context, we mention the work of Ref.~\cite{Kirscher:2018dwo}, which stated that pairs of a deep ground state and a shallow excited
state with the same quantum numbers as in $^4$He also occur in larger nuclei like $^{12}$C and $^{16}$O. Consequently, it is worth mentioning that within NLEFT, the first {\em ab initio} calculation of the Hoyle state
in $^{12}$C was performed~\cite{Epelbaum:2011md}, which, together with the ground state, is the arguably the most known of such pairs.
{ Taking these achievements into account, we
 believe that the NLEFT framework  is well suited to address the issue of the $\alpha$-particle transition form factor.}

\section{Formalism}

In Ref.~\cite{Lu:2018bat} a  minimal nuclear interaction was
constructed that
reproduces the ground state properties of light nuclei, medium-mass nuclei, and neutron matter simultaneously
with no more than a few percent error in the energies and charge radii.
It is given by the SU(4)-invariant leading-order effective field theory
without pions, formulated on a periodic cubic box of $L^3$. The Hamiltonian reads
\begin{equation}
H_{{\rm SU(4)}}=H_{\rm free}+\frac{1}{2!}C_{2}\sum_{\bm{n}}\tilde{\rho}(\bm{n})^{2}
+\frac{1}{3!}C_{3}\sum_{\bm{n}}\tilde{\rho}(\bm{n})^{3},\label{eq:HSU4}
\end{equation}
where  $\bm{n}=(n_{x,}n_{y},n_{z})$ are the lattice coordinates,
$H_{\rm free}$ is the free nucleon Hamiltonian with nucleon mass $m=938.9$~MeV.
The lattice spacing is $a = 1.32$~fm, which corresponds to a momentum cutoff $\Lambda = \pi/a\simeq 471\,$MeV,
which is also the optimal resolution scale to unravel the hidden spin-isospin symmetry of QCD in the limit of
a large number of colors \cite{Lee:2020esp}.
 The density operator $\tilde{\rho}(\bm{n})$ is defined in the same manner
as introduced in Ref.~\cite{Elhatisari:2017eno},
\begin{equation}
\tilde{\rho}(\bm{n})=\sum_{i}\tilde{a}_{i}^{\dagger}(\bm{n})\tilde{a}_{i}(\bm{n})
+s_{L}\sum_{|\bm{n}^{\prime}-\bm{n}|=1}\sum_{i}\tilde{a}_{i}^{\dagger}(\bm{n}^{\prime})\tilde{a}_{i}(\bm{n}^{\prime}),
\end{equation}
where $i$ is the joint spin-isospin index, $s_{L}$ is the local smearing parameter, and the nonlocally smeared annihilation and
creation operators with parameter $s_{NL}$ are defined as
\begin{equation}
\tilde{a}_{i}(\bm{n})=a_{i}(\bm{n})+s_{NL}\sum_{|\bm{n}^{\prime}-\bm{n}|=1}a_{i}(\bm{n}^{\prime}).
\end{equation}
The summation over the spin and isospin implies that the interaction is SU(4)
invariant. The parameter $s_L$ controls the strength of the local part of the
interaction, while $s_{NL}$ controls the strength of the nonlocal part of
the interaction.
The parameters $C_{2}$ and $C_{3}$ give the overall strength of the two-body
and three-body interactions, respectively. For a given value of$s_{NL}$,  $C_{2}$,  $C_{3}$ and $s_{L}$  are determined by fitting to  $A \le 3$ data. Then, the optimal strength and range of the local and non-local parts of the interactions are defined by parameterizing the nuclear binding energies with nuclei with $A\geq 16$ with the Bethe-Weizs{\"a}cker mass formula.
Note that the local part the interactions is an important factor in nuclear binding, especially for
the  $\alpha$-$\alpha$ interaction~\cite{Elhatisari:2016owd}. These parameters have been
determined in Ref.~\cite{Lu:2018bat} as $C_2 = -3.41 \cdot 10^{-7}\,$MeV$^{-2}$, 
$C_3 = -1.4 \cdot 10^{-14}\,$MeV$^{-5}$, $s_{NL} = 0.5$, and $s_L = 0.061$,
and they  will be used throughout this work. The effects from the Coulomb interaction are
included in perturbation theory. For details, see Ref.~\cite{Lu:2018bat}.

The transition form factor $F(q)$ of the monopole transition is related to the transition density $\rho_{\rm tr}(r)$
by 
\begin{eqnarray}
F(q) &=& \frac{4\pi}{Z}\int_0^\infty \rho_{\rm tr}(r) j_0(qr)r^2dr 
\nonumber\\
&=& \frac{1}{Z}\sum_{\lambda=1}^\infty 
\frac{(-1)^\lambda}{(2\lambda+1)!} q^{2\lambda} \langle r^{2\lambda}\rangle_{\rm tr}~,
\end{eqnarray}
with $Z$ the charge of the nucleus under consideration. Here $Z=2$, and $\rho_{\rm
tr}(r) = \langle 0_1^+ | \hat\rho(\vec{r})|0_2^+\rangle$ is the matrix element of the charge density
operator $\hat\rho(\vec{r})$ between the ground state $0_1^+$ and the first excited $0_2^+$ state.
This definition differs from the one used in Ref.~\cite{Kegel:2021jrh} by a factor of $Z/\sqrt{4\pi}$. It is also
interesting to consider the low-$q$ expansion of the transition form factor. We use the definition of Ref.~\cite{Kegel:2021jrh},
\begin{equation}\label{eq:lowq}
\frac{Z|F(q^2)|}{q^2} = \frac{1}{6}\, \langle r^2\rangle_{\rm tr}\,\left[ 1 - \frac{q^2}{20}{\cal R}^2_{\rm tr} + {\cal O}(q^4)\right]~,
\end{equation}
with ${\cal R}^2_{\rm tr} = \langle r^4\rangle_{\rm tr}/\langle r^2\rangle_{\rm tr}$. The corresponding parameters were
extracted in Ref.~\cite{Kegel:2021jrh} as $\langle r^2\rangle_{\rm tr} = 1.53\pm 0.05\,$fm$^2$ and ${\cal R}_{\rm tr}  = 4.56\pm 0.15\,$fm.

\section{Results and discussion}

The first excited state of $^4$He is a resonance that sits just above the $^3$H$+p$ threshold.  In order to study  this continuum state, we perform calculations using three different cubic periodic boxes with lengths $L=10, 11, 12$ in lattice units, corresponding to $L=13.2$~fm, $14.5$~fm, $15.7$~fm.
We then compare results for the different box sizes in order to quantify the residual uncertainties in the resonance energy and wave function due to the finite volume and decay width. The lattice calculations performed in this work follow the same methods as presented in Ref.~\cite{Shen:2022bak} for the low-lying states of $^{12}$C.  We use the Euclidean time projection operator $\exp(-Ht)$ to prepare the low-lying states of $^4$He, starting from some set of initial states with the desired quantum numbers.  The operator $\exp(-Ht)$ is implemented using auxiliary-field Monte Carlo simulations with time step size $a_t = (1000\,{\rm MeV})^{-1} = 0.197\,{\rm fm}^{-1}$.  The total number of time steps is denoted $L_t$, and so $t = L_t a_t$.  While we do not compute the decay width of the $0^+_2$ state in this work, new computational algorithms for computing widths of resonances from finite-volume lattice Monte Carlo simulations are currently under development. 

We perform coupled channel calculations using three different initial states composed of shell-model wave functions. The first channel contains four particles in
the $1s_{1/2}$ state with oscillator frequency $\hbar\omega = 
20\,$MeV.  The second channel has all three particles for $^3$H in the $1s_{1/2}$ state and an excited proton in the $2s_{1/2}$ state.
The third channel  has one neutron in the $2s_{1/2}$ state and the remaining nucleons in the $1s_{1/2}$ state. In these channels, we use  $\hbar\omega = 6\,$MeV for the initial state of the excited particle and  $\hbar\omega = 14\,$MeV for the initial state of the
$3N$ system. The three-channel calculation accelerates the exponential convergence of the two lowest lying $0^+$ states in the limit of infinite Euclidean time, $L_t\to\infty$. {In fact, the values of $\hbar\omega$ in the second and third channel were tuned
to optimize this convergence, however, the final results do not depend on these particular choices, see \cite{SM} for details.}
The corresponding ground and first excited state energies are collected in Tab.~\ref{tab:1}.
These compare well with the experimental values of $E(0_1^+) =-28.30\,$MeV and $E(0_2^+) =-8.09\,$MeV.
\begin{table}[h]
\begin{tabular}{|c|c|c|c|}
\hline
$L$ [fm] & $E(0_1^+)$ [MeV] &  $E(0_2^+)$ [MeV] & $\Delta E$ [MeV] \\
\hline
$13.2$  &  $-28.32(3)$  & $ -8.37(14)$ & $0.28(14)$ \\
$14.5$  &  $-28.30(3)$  & $ -8.02(14)$ & $0.42(14)$ \\
$15.7$  &  $-28.30(3)$  & $ -7.96(9)$ & {$0.40(9)$} \\
\hline
\end{tabular}
\caption{Energy of the $^4$He ground state $(0_1^+)$ and the first excited state $(0_2^+)$  for
different box sizes $L$. Here, $\Delta E = E(0_2^+)-E(^3{\rm H})$ for the same box length $L$. The error bars include stochastic errors
and uncertainties in the Euclidean time extrapolation.
\label{tab:1}}
\end{table}
The ground state energies of the $3N$ systems using an exact calculation at $L=12$ are $E(^3{\rm H}) = -8.36\,$MeV and  
$E(^3{\rm He}) = -7.65\,$MeV, which compare well with the experimental values of $-8.48$ and $-7.72\,$MeV, 
respectively. We note some difference to the energies given in Ref.~\cite{Lu:2018bat}, which can be traced back to the use
of larger volumes here. However, for the cases of $L=8,9,10,$ our results align with those presented in Ref.~\cite{Lu:2018bat}, albeit with slightly enlarged uncertainties. {We have performed calculations with the Coulomb interaction treated
non-perturbatively, the resulting energy differences are below 1 per mille, see \cite{SM}.}
For $L=12$, we find $\Delta E=E(0_2^+)-E(^3{\rm H})= {0.40(9)}\,$MeV, consistent with the main finding of Ref.~\cite{Michel:2023ley}. We note that a recent paper finds that $E(0_2^+)$ and $E(^3{\rm H})$ are much closer together \cite{Gattobigio:2023fmo}, in contrast with experimental observations. The results for $L=10$ and $L=11$ give very similar results for $\Delta E$.  We also find only small differences in the transition form factor results for $L = 10, 11, 12$. In the following we represent results for $L=12$, corresponding to a box volume of $V= (15.7\,{\rm fm})^3$.  {Note further that when switching off the
3NF, the value of $\Delta E$ and the transition form factor increase, entirely consistent with the findings of
Ref.~\cite{Michel:2023ley}, see \cite{SM} for details.} 

Since we are using a short-ranged SU(4) interaction, we can resort to the Efimov analysis of
Refs.~\cite{Hammer:2006ct,von2009signatures}
to give an additional argument in favor of explaining why the $0_2^+$ state can be accurately predicted by NLEFT using a minimal nuclear interaction. We employ the universal Efimov
tetramer relation between the second state in $^4$He and the triton energy, $E(0_2^+) = 1.01 \times E(^3{\rm H})$. In the absence of the Coulomb interaction, the $0_2^+$ state should be $0.01 \times 8.48 = 0.084\,$MeV below the triton threshold.
The Coulomb energy has no {effect} on the triton.  Adding the Coulomb energy for $0_2^+$ from our lattice calculations with $L=12$, we get $0.46(2)-0.08=0.38(2)\,$MeV,  which is very close to the observed value of $0.40$~MeV.

\begin{figure}[!htbp]
  \centering
  \includegraphics[width=0.50\textwidth]{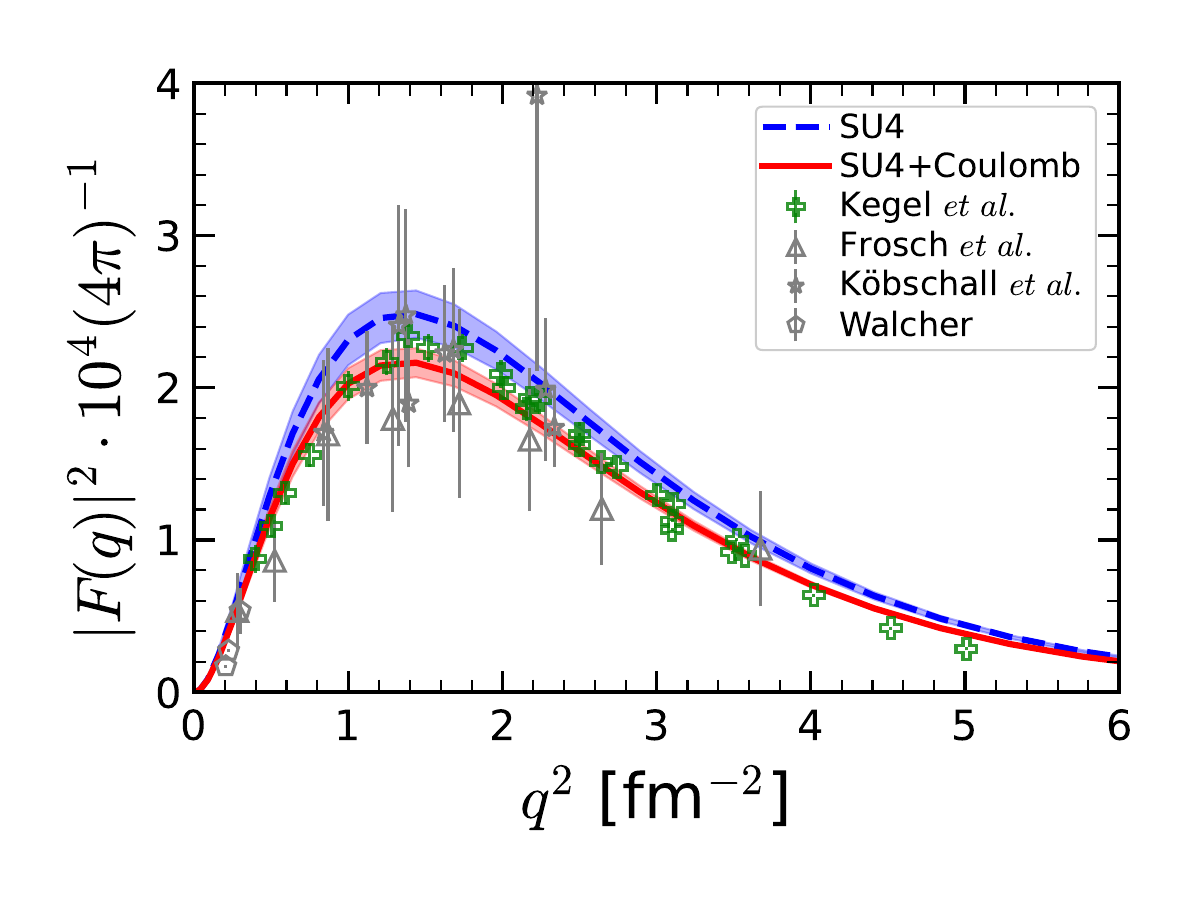}\\
  \caption{Calculated monople form factor of the $0_2^+\to 0_1^+$ transition in $^4$He compared to the
   recent data from Mainz~\cite{Kegel:2021jrh} (green squares) and the older data from 
   Refs.~\cite{Frosch:1968sns,Walcher:1970vkv, Kobschall:1983na} (grey symbols).   Blue dashed line:
   SU(4) symmetric strong interaction with all parameters determined in Ref.~\cite{Lu:2018bat}. Red solid line:
   adding the Coulomb interaction perturbatively. The uncertainty bands in the lattice results include stochastic errors
and uncertainties in the Euclidean time extrapolation.}    
  \label{fig1}
\end{figure}

Next, we turn to the analysis of the transition form factor, denoted as $F(q)$. In the framework of NLEFT, observables such as nucleon density distributions, charge radii and form factors can be computed using the pinhole algorithm~\cite{Elhatisari:2017eno}, which performs a Monte Carlo sampling of the $A$-body density of the nucleus in position space. Furthermore, the pinhole algorithm can be combined with the first order perturbation theory to compute the corrections to these observables~\cite{Lu:2018bat}. In this work, we compute the transition form factor $F(q)$ using the pinhole algorithm while the Coulomb interaction is treated using perturbation theory.  Further details and additional calculations extending this work will be presented in  Ref.~\cite{tFF}. First, we consider the SU(4)-symmetric interactions without Coulomb. The resulting 
form factor is depicted by the blue dashed line in Fig.~\ref{fig1}. It somewhat overshoots the data,
although the error band associated with stochastic errors and the large $L_t$ extrapolation almost encompasses the data.
Including the Coulomb interaction leads to an overall reduction of the transition form factor as shown by
the red solid line in  Fig.~\ref{fig1}. Overall, we achieve a good reproduction of the data and the
uncertainty band is also somewhat reduced. This is due to the fact that inclusion of the Coulomb interaction
leads to smaller fluctuations in the Monte Carlo data when extrapolating to large $L_t$.
Consequently, we find that the  nuclear interaction defined in Ref.~\cite{Lu:2018bat}, which has already been shown to reproduce the essential elements of nuclear
binding, also leads
to a good description of the $\alpha$-particle transition $0_2^+\to 0_1^+$  form factor without adjusting any parameters.
{ Note that we have also analyzed the many-body uncertainty underlying our minimal interaction, as detailed
  in \cite{SM}. For $A\leq 4$, this error is much smaller than the statistical errors from the MC simulation and the
  $L_t$ extrapolation.}

\begin{figure}[!htbp]
  \centering
  \includegraphics[width=0.450\textwidth]{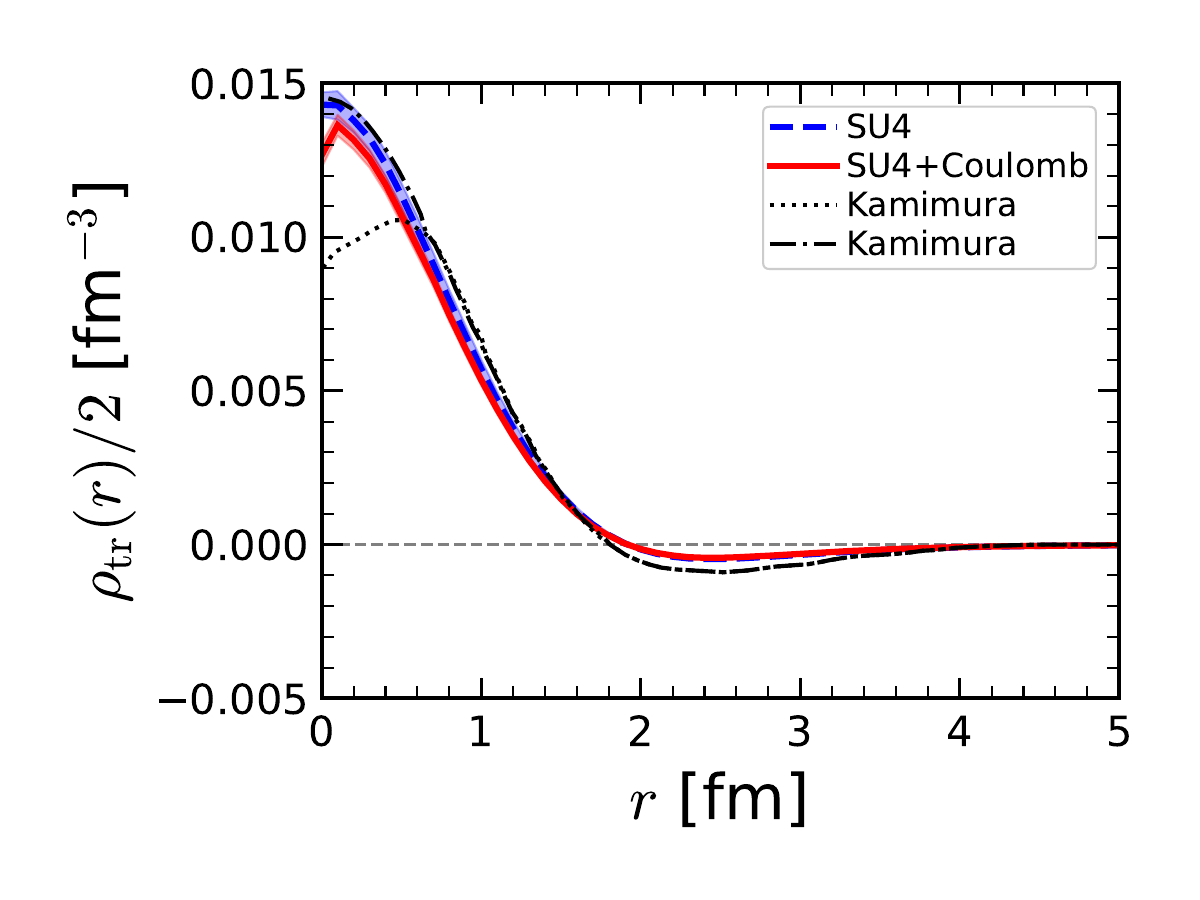}\\
  \caption{The transition charge density for the SU(4) interaction (blue dashed line) and the SU(4) plus Coulomb 
    interaction (red solid line) in comparison to the results of Kamimura in Ref.~\cite{Kamimura:2023tvl}.  
    The black dotted (dash-dotted ) line correponds to no node (a node in the transition form factor near $q^2 = 14~{\rm fm}^{-2}$.)
     The uncertainty bands in the lattice results include stochastic errors and uncertainties in the Euclidean time extrapolation. }    
  \label{fig2}
\end{figure}

The transition density $\rho_{\rm tr}(r)$  underlying the form factor is displayed in Fig.~\ref{fig2}, for the
SU(4) interaction and the inclusion of the Coulomb interaction.  The corresponding curves are very similar to the results of
Ref.~\cite{Kamimura:2023tvl}, though we find a less pronounced central depletion when the Coulomb force
is included.
Note, however, that the definition used there is  based on averaging over all four  nucleons, while our definition is the 
charge (proton) density. We account for the charge radius of the proton, $r_p =0.84\,$fm~\cite{Lin:2021xrc},
while in Ref.~\cite{Kamimura:2023tvl} a more phenomenological proton size-factor is used.

\begin{figure}[!htbp]
  \centering
  \includegraphics[width=0.450\textwidth]{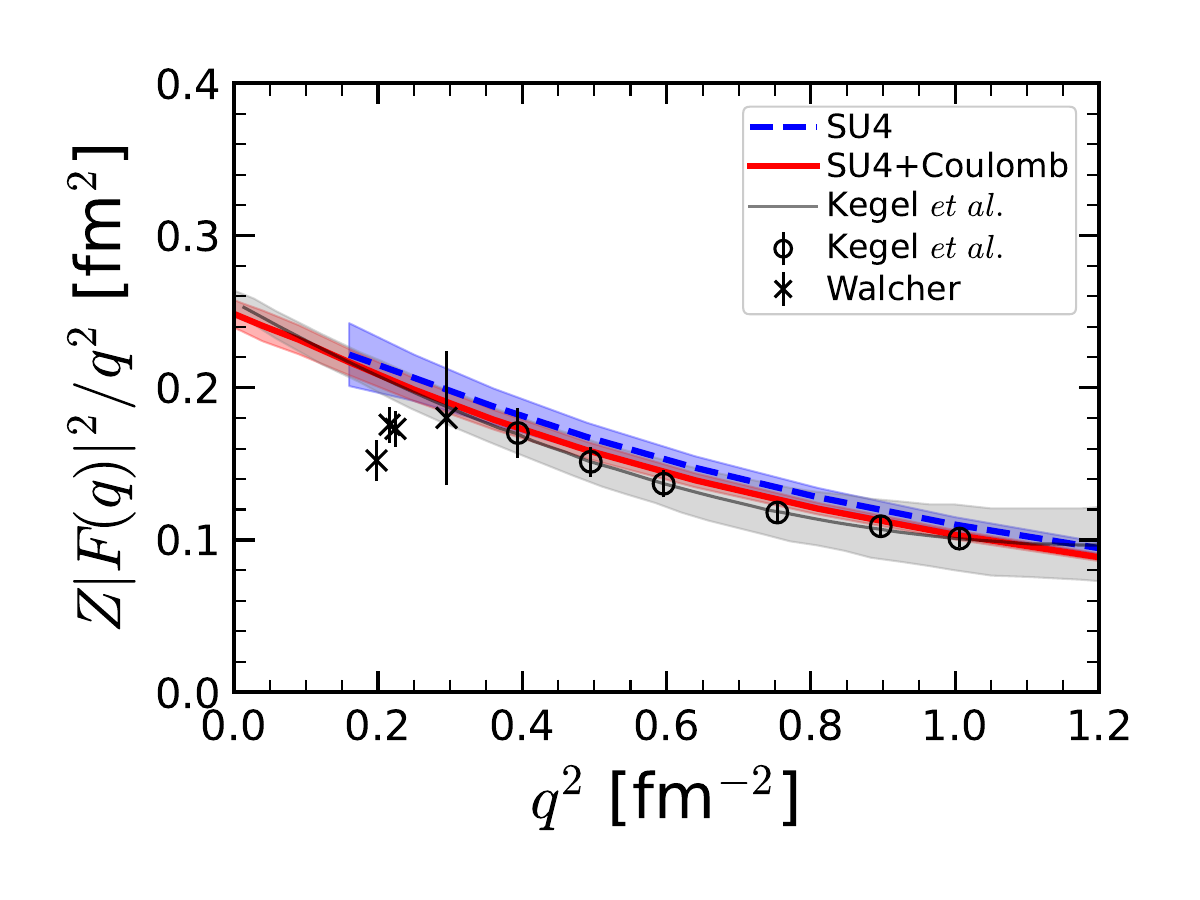}\\
  \caption{The low-momentum expansion of the transition form factor for the SU(4) interaction (blue dashed line)
    and the SU(4) plus Coulomb 
    interaction (red solid line) in comparison to the results of Refs.~\cite{Kegel:2021jrh,Walcher:1970vkv}.
     The uncertainty bands for the NLEFT results are due to the $L_t$ extrapolation. The grey band is from 
     Ref.~\cite{Kegel:2021jrh}.}    
  \label{fig3}
\end{figure}

Now, we consider the low-momentum expansion as given in Eq.~\eqref{eq:lowq}. The resulting
curves for the SU(4) and SU(4) plus Coulomb interactions are shown in Fig.~\ref{fig3}. 
Our results are in good agreement with the results of Ref.~\cite{Kegel:2021jrh}, which is shown by the grey band. 
We note again that the error band of the NLEFT calculation is reduced when the Coulomb interaction
is included. The corresponding moments of the low-$q$ expansion are
$\langle r^2\rangle_{\rm tr} = 1.48(1)~{\rm fm}^2$ and ${\cal R}_{\rm tr} = 3.61(3)~{\rm fm}$ for the SU(4)
interaction fitted in the range $q^2=0.09-0.49\,$fm$^{-2}$, and
$\langle r^2\rangle_{\rm tr} = 1.49(1)~{\rm fm}^2$ and ${\cal R}_{\rm tr} = 4.00(4)~{\rm fm}$
for the SU(4) plus Coulomb interaction, fitted in the  range $q^2=0.04-0.25\,$fm$^{-2}$, as the signals
are less noisy at low $q^2$ when the Coulomb interaction is included.

\section{Summary and discussion}

In this letter, we have used a minimal nuclear interaction that allows to describe the gross features of nuclei
and nuclear matter with no more than a few percent error to postdict the alpha particle transition
form factor from the first excited to the ground state. This interaction accounts for SU(4) symmetric two-
and three-body terms as well as the Coulomb interaction with only four parameters that previously had been
determined in Ref.~\cite{Lu:2018bat}.  Firstly, we reproduce the energies of the
ground state and the first excited state of $^4$He.  This is a known prerequisite to properly describe the
form factor due to the closeness of the first excited state to the $^3$H+p threshold~\cite{Hiyama:2004nf,Michel:2023ley}. Having met that prerequisite, we find that the description of the transition form factor and its low-energy expansion is quite satisfactory. The nuclear forces relevant to this system are under good control, and we do not find the puzzle
mentioned in Ref.~\cite{Kegel:2021jrh}.  
We were able to accurately reproduce the position of the energy $0^+_2$ relative to $^3$H with a simple interaction and no parameter tuning.  This strongly suggests a link between the tuning of the $\alpha$-$\alpha$ interaction already performed in Ref.~\cite{Lu:2018bat} and the required tuning of the $p$-$^3$H interaction to get the correct energy of $0^+_2$ relative
to $^3$H.  In the future, calculations of the monopole transition form factor can be made more systematic and accurate by using high-fidelity chiral interactions and the machinery of wave function matching \cite{Elhatisari:2022qfr}.


\section{acknowledgments}
We are grateful for discussion with the members of the NLEFT Collaboration.  In particular, we acknowledge the work of Bing-Nan Lu who developed the interaction in Ref.~\cite{Lu:2018bat}.  This work is supported in part by the European
Research Council (ERC) under the European Union's Horizon 2020 research
and innovation programme (ERC AdG EXOTIC, grant agreement No. 101018170),
by DFG and NSFC through funds provided to the
Sino-German CRC 110 ``Symmetries and the Emergence of Structure in QCD" ({DFG project-ID 196253076 - TRR 110,
NSFC grant No. 12070131001}).
The work of UGM was supported in part by VolkswagenStiftung (Grant no. 93562)
and by the CAS President's International
Fellowship Initiative (PIFI) (Grant No.~2018DM0034).
The work of SE is supported in part by the Scientific and Technological Research Council of Turkey (TUBITAK project no. 120F341).
The work of DL is supported in part by the U.S. Department of Energy (Grant
Nos. DE-SC0021152, DE-SC0013365, DE-SC0023658) and the Nuclear Computational Low-Energy Initiative (NUCLEI) SciDAC project.
The authors gratefully acknowledge the Gauss Centre for Supercomputing e.V. (www.gauss-centre.eu)
for funding this project by providing computing time on the GCS Supercomputer JUWELS
at J\"ulich Supercomputing Centre (JSC).


\newpage
\begin{appendix}
\begin{onecolumngrid}

\renewcommand{\thefigure}{S\arabic{figure}}
\setcounter{figure}{0}

\section*{Supplemental Material}

\subsection{Dependence on the shell-model initial states}

Here, we discuss the dependence of the results of the first excited state for our three coupled channels
calculation on $\hbar\omega$ in the second and third channel. We keep the value of  $\hbar\omega$ in
the first channel fixed, because this gives the ground state which has little influence on the
first excited state. In Fig.~\ref{figS1}, we show the result for the first excited state for
various combinations of $\hbar \omega_1$ and $\hbar \omega_2$, corresponding to the
$3N$ system and excited particle. Although the starting energies vary a lot, the final result is entirely stable.

\begin{figure}[!htbp]
  \centering
  \includegraphics[width=0.450\textwidth]{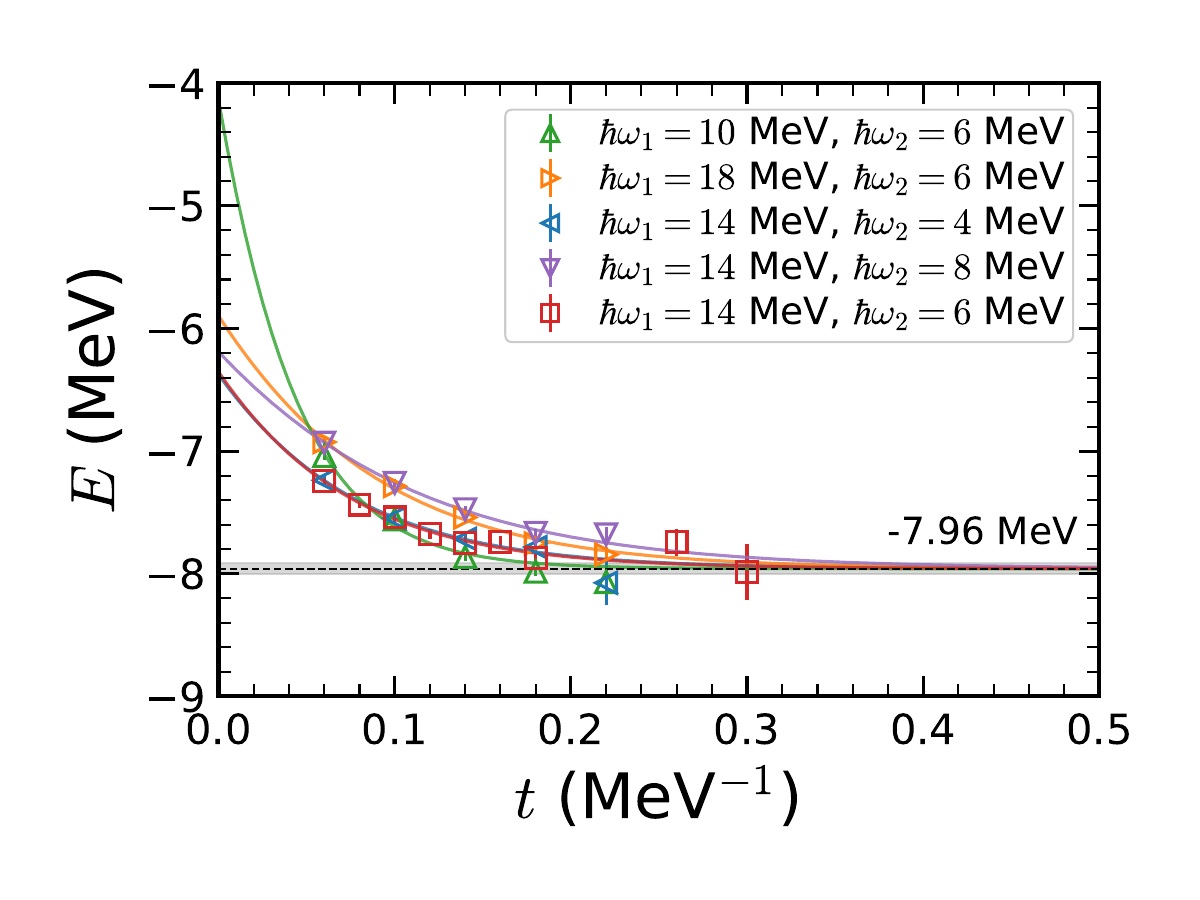}\\
  \caption{Calculation of the energy of the first excited state by varying  the
    $\hbar\omega$ values in the second and third channel.
    }
  \label{figS1}
\end{figure}

\subsection{Influence of the three-body forces}

Here, we address the issue of the three-nucleon force (3NF) contribution to the energies and
the transition form factor. If we switch off the 3NF, $\Delta E$ increases to $\Delta E = 0.50(6)$~MeV
and the form factor comes out above the data, see Fig.~S2, and Fig.~S3 for the low-momentum expansion.
This is completely consistent with the
findings of Michel et al. in  Ref.~\cite{Michel:2023ley}.

\begin{figure}[!h]
  \centering
  \includegraphics[width=0.450\textwidth]{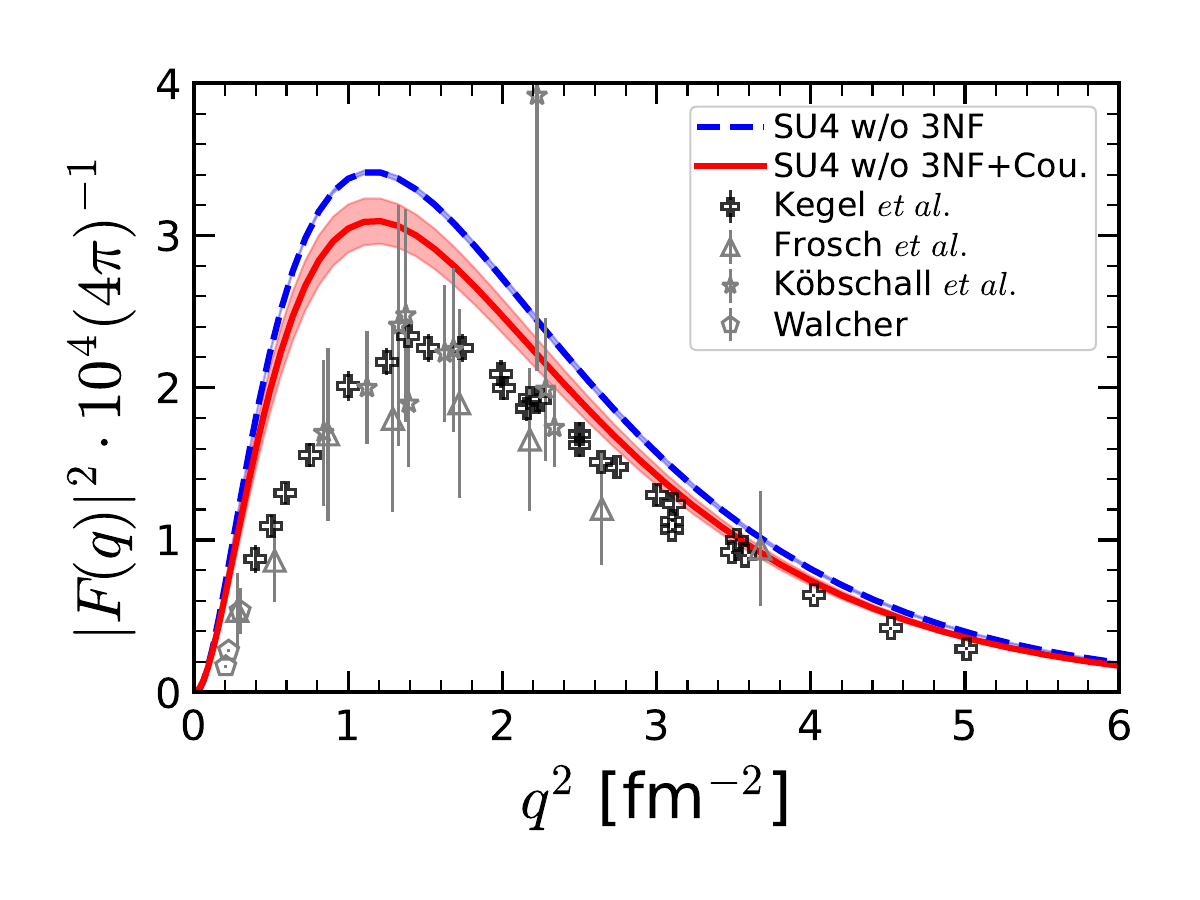}\\
  \caption{Calculated monople form factor of the $0_2^+\to 0_1^+$ transition in $^4$He
    with the 3NF contribution switched off compared to the
   recent data from Mainz~\cite{Kegel:2021jrh} (green squares) and the older data from 
   Refs.~\cite{Frosch:1968sns,Walcher:1970vkv, Kobschall:1983na} (grey symbols).   Blue dashed line:
   SU(4) symmetric strong interaction with all parameters determined in Ref.~\cite{Lu:2018bat}. Red solid line:
   adding the Coulomb interaction perturbatively. The uncertainty bands in the lattice results include stochastic errors
and uncertainties in the Euclidean time extrapolation.}    
  \label{figS2}
\end{figure}

\begin{figure}[!ht]
  \centering
  \includegraphics[width=0.450\textwidth]{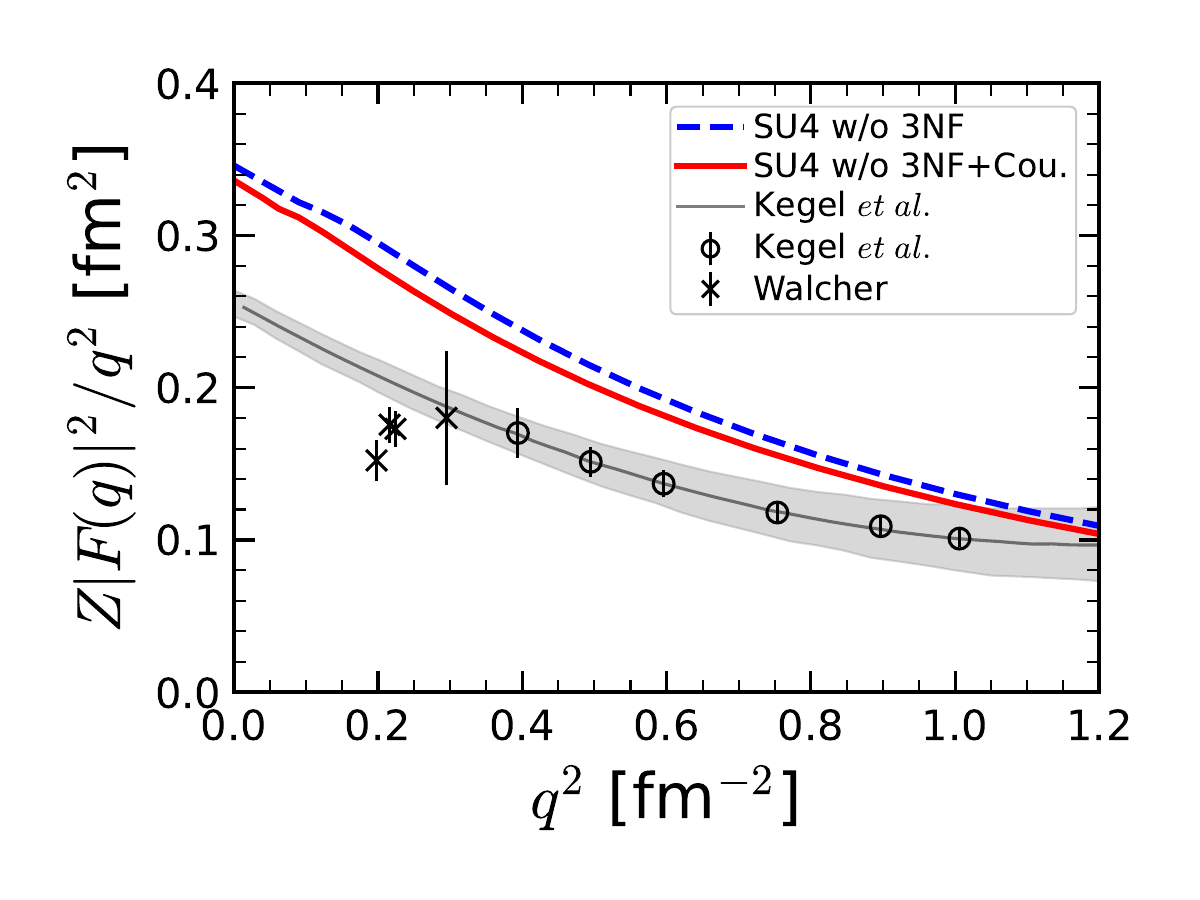}\\
  \caption{The low-momentum expansion of the transition form factor for the SU(4) interaction
    with the 3NF contribution switched off (blue dashed line)
    and the SU(4) plus Coulomb 
    interaction (red solid line) in comparison to the results of Refs.~\cite{Kegel:2021jrh,Walcher:1970vkv}.
     The uncertainty bands for the NLEFT results are due to the $L_t$ extrapolation. The grey band is from 
     Ref.~\cite{Kegel:2021jrh}.}    
  \label{figS3}
\end{figure}

\subsection{Many-body uncertainties}

In Ref.~\cite{Lu:2019nbg}, many-body observables were used to pin down the non-local
smearing parameter $s_{NL}$ by calculating the liquid drop constants $a_V$ and $a_S$
in the mass range $16\leq A\leq 40$. This led to the preferred value of $s_{NL}=0.5$.
However, in that paper (see Fig.~S3 therein) $a_V$ and $a_S$ were also calculated for the
range of $0.4 \leq s_{NL} \leq 0.6$. This dependence on $s_{NL}$ can indeed be used
to quantify the many-body uncertainty of the minimal interaction in that given mass
range. This is shown in Fig.~\ref{figS4}, where we display the strong interaction contribution
of the nuclear masses normalized to the reference value of $s_{NL}=0.5$.
Extrapolating this down to the $^4$He nucleus, this type of uncertainty is well below
one percent and thus completely negligible compared to the errors for the large $L_t$ extrapolation and the
stochastic errors.

\begin{figure}[!htbp]
\centering
\includegraphics[width=0.450\textwidth]{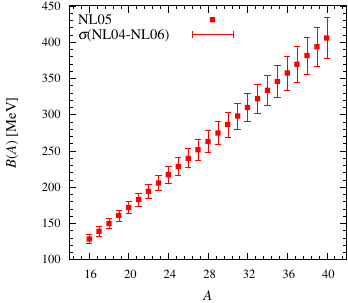}\\
\caption{Nuclear masses (without Coulomb interaction) for varying
non-local
smearing normalized to the value with $s_{NL}=0.5$ in the range
$16\leq A\leq 40$.}
\label{figS4}
\end{figure}

\subsection{Coulomb interaction: Perturbative and non-perturbative treatment}

In the main text, we have considered the effects of the Coulomb interaction in first
order perturbation theory to be consistent with Ref.~~\cite{Lu:2019nbg}. One can,
however, include the Coulomb interaction also non-perturbatively. The simplest
nucleus of relevance here is $^3$He.  Upon exact diagonalization in a box with $L=12$, we find $E(^3{\rm He}) = -7.654$~MeV (perturbative Coulomb) $E(^3{\rm He}) = -7.660$~ MeV (non-perturbative Coulomb).  This difference is well below 1~per~mille, and we expect similar agreement between perturbative and non-perturbative Coulomb for $^4$He.

\end{onecolumngrid}
\end{appendix}


\begin{thebibliography}{99}

\bibitem{Kamada:2001tv}
H.~Kamada, A.~Nogga, W.~Gloeckle, E.~Hiyama, M.~Kamimura, K.~Varga, Y.~Suzuki, M.~Viviani, A.~Kievsky and S.~Rosati, \textit{et al.}
Phys. Rev. C \textbf{64} (2001), 044001
[arXiv:nucl-th/0104057 [nucl-th]].

\bibitem{Bacca:2012xv}
S.~Bacca, N.~Barnea, W.~Leidemann and G.~Orlandini,
Phys. Rev. Lett. \textbf{110} (2013) no.4, 042503
[arXiv:1210.7255 [nucl-th]].

\bibitem{Kirscher:2018dwo}
J.~Kirscher and H.~W.~Grie\ss{}hammer,
Eur. Phys. J. A \textbf{54} (2018) no.8, 137
[arXiv:1803.05949 [nucl-th]].

\bibitem{Kegel:2021jrh}
  S.~Kegel, P.~Achenbach, S.~Bacca, N.~Barnea, J.~Bericic, D.~Bosnar, L.~Correa, M.~O.~Distler, A.~Esser and H.~Fonvieille,
  \textit{et al.}
Phys. Rev. Lett. \textbf{130} (2023) no.15, 152502
[arXiv:2112.10582 [nucl-ex]].


\bibitem{Efros:1994iq}
V.~D.~Efros, W.~Leidemann and G.~Orlandini,
Phys. Lett. B \textbf{338} (1994), 130-133
[arXiv:nucl-th/9409004 [nucl-th]].



\bibitem{Beane:2000fx}
S.~R.~Beane, P.~F.~Bedaque, W.~C.~Haxton, D.~R.~Phillips and M.~J.~Savage,
[arXiv:nucl-th/0008064 [nucl-th]].

\bibitem{Epelbaum:2008ga}
E.~Epelbaum, H.~W.~Hammer and U.-G.~Mei{\ss}ner,
Rev. Mod. Phys. \textbf{81} (2009), 1773-1825
[arXiv:0811.1338 [nucl-th]].

\bibitem{Machleidt:2011zz}
R.~Machleidt and D.~R.~Entem,
Phys. Rept. \textbf{503} (2011), 1-75
[arXiv:1105.2919 [nucl-th]].

\bibitem{Hammer:2019poc}
H.~W.~Hammer, S.~K\"onig and U.~van Kolck,
Rev. Mod. Phys. \textbf{92} (2020) no.2, 025004
[arXiv:1906.12122 [nucl-th]].

\bibitem{Epelbaum:2019kcf}
E.~Epelbaum, H.~Krebs and P.~Reinert,
Front. in Phys. \textbf{8} (2020), 98
[arXiv:1911.11875 [nucl-th]].


\bibitem{EE}
E.~Epelbaum,
Physics {\bf 16} (2023) 58.

\bibitem{Michel:2023ley}
N.~Michel, W.~Nazarewicz and M.~P\l{}oszajczak,
{\em Phys. Rev. Lett.} (2023) in print
[arXiv:2306.05192 [nucl-th]].

\bibitem{Hiyama:2004nf}
E.~Hiyama, B.~F.~Gibson and M.~Kamimura,
Phys. Rev. C \textbf{70} (2004), 031001

\bibitem{Kamimura:2023tvl}
M.~Kamimura,
PTEP \textbf{2023} (2023) no.7, 071D01
[arXiv:2306.07268 [nucl-th]].

\bibitem{Lee:2008fa}
D.~Lee,
Prog. Part. Nucl. Phys. \textbf{63} (2009), 117-154
[arXiv:0804.3501 [nucl-th]].

\bibitem{Lahde:2019npb}
T.~A.~L\"ahde and U.-G.~Mei\ss{}ner,
Lect. Notes Phys. \textbf{957} (2019), 1-396
Springer, 2019,
ISBN 978-3-030-14187-5, 978-3-030-14189-9

\bibitem{Lu:2018bat}
B.~N.~Lu, N.~Li, S.~Elhatisari, D.~Lee, E.~Epelbaum and U.-G.~Mei\ss{}ner,
Phys. Lett. B \textbf{797} (2019), 134863
[arXiv:1812.10928 [nucl-th]].

\bibitem{Lu:2019nbg}
B.~N.~Lu, N.~Li, S.~Elhatisari, D.~Lee, J.~E.~Drut, T.~A.~L\"ahde, E.~Epelbaum and U.-G.~Mei\ss{}ner,
Phys. Rev. Lett. \textbf{125} (2020) no.19, 192502
[arXiv:1912.05105 [nucl-th]].

\bibitem{Ren:2023ued}
Z.~Ren, S.~Elhatisari, T.~A.~L\"ahde, D.~Lee and U.-G.~Mei\ss{}ner,
[arXiv:2305.15037 [nucl-th]].

\bibitem{Shen:2022bak}
S.~Shen, S.~Elhatisari, T.~A.~L\"ahde, D.~Lee, B.~N.~Lu and U.-G.~Mei\ss{}ner,
Nature Commun. \textbf{14} (2023) no.1, 2777
[arXiv:2202.13596 [nucl-th]].

\bibitem{Chernykh:2007zz}
M.~Chernykh, H.~Feldmeier, T.~Neff, P.~von Neumann-Cosel and A.~Richter,
Phys. Rev. Lett. \textbf{98} (2007), 032501

\bibitem{Epelbaum:2011md}
E.~Epelbaum, H.~Krebs, D.~Lee and U.-G.~Mei{\ss}ner,
Phys. Rev. Lett. \textbf{106} (2011), 192501
[arXiv:1101.2547 [nucl-th]].

\bibitem{Lee:2020esp}
D.~Lee, S.~Bogner, B.~A.~Brown, S.~Elhatisari, E.~Epelbaum, H.~Hergert, M.~Hjorth-Jensen, H.~Krebs, N.~Li and B.~N.~Lu, \textit{et al.}
Phys. Rev. Lett. \textbf{127} (2021) no.6, 062501
[arXiv:2010.09420 [nucl-th]].

\bibitem{Elhatisari:2016owd}
S.~Elhatisari, N.~Li, A.~Rokash, J.~M.~Alarc\'on, D.~Du, N.~Klein, B.~n.~Lu, U.~G.~Mei\ss{}ner, E.~Epelbaum and H.~Krebs, \textit{et al.}
Phys. Rev. Lett. \textbf{117}, no.13, 132501 (2016)
[arXiv:1602.04539 [nucl-th]].


\bibitem{SM} see the {\em Supplemental Material}.

\bibitem{Gattobigio:2023fmo}
M.~Gattobigio and A.~Kievsky,
Few Body Syst. \textbf{64}, no.4, 86 (2023)
[arXiv:2305.16814 [nucl-th]].


\bibitem{Hammer:2006ct}
H.~W.~Hammer and L.~Platter,
Eur. Phys. J. A \textbf{32} (2007), 113-120
[arXiv:nucl-th/0610105 [nucl-th]].

\bibitem{von2009signatures}
J.~von~Stecher, J.~P.~D'Incao and C.~H.~Greene,  
Nature Physics {\bf 5} (6) (2009) 417-421

\bibitem{Frosch:1968sns}
R.~F.~Frosch, R.~E.~Rand, H.~Crannell, J.~S.~McCarthy, L.~R.~Suelzle and M.~R.~Yearian,
Nucl. Phys. A \textbf{110} (1968), 657-673

\bibitem{Walcher:1970vkv}
T.~Walcher,
Phys. Lett. B \textbf{31} (1970), 442-444

\bibitem{Kobschall:1983na}
G.~Kobschall, C.~Ottermann, K.~Maurer, K.~Rohrich, C.~Schmitt and V.~H.~Walther,
Nucl. Phys. A \textbf{405} (1983), 648-652


\bibitem{Elhatisari:2017eno}
S.~Elhatisari, E.~Epelbaum, H.~Krebs, T.~A.~L\"ahde, D.~Lee, N.~Li, B.~n.~Lu, U.-G.~Mei\ss{}ner and G.~Rupak,
Phys. Rev. Lett. \textbf{119} (2017) no.22, 222505
[arXiv:1702.05177 [nucl-th]].

\bibitem{tFF}
S.~Shen, S.~Elhatisari, D.~Lee and U.-G.~Mei{\ss}ner, {\em in preparation}.


\bibitem{Lin:2021xrc}
Y.~H.~Lin, H.~W.~Hammer and U.-G.~Mei\ss{}ner,
Phys. Rev. Lett. \textbf{128} (2022) no.5, 052002
[arXiv:2109.12961 [hep-ph]].

\bibitem{Elhatisari:2022qfr}
S.~Elhatisari, L.~Bovermann, E.~Epelbaum, D.~Frame, F.~Hildenbrand, M.~Kim, Y.~Kim, H.~Krebs, T.~A.~L\"ahde and D.~Lee, \textit{et al.}
[arXiv:2210.17488 [nucl-th]].

\end{thebibliography}

\begin{thebibliography}{99}

\bibitem{Michel:2023ley}
N.~Michel, W.~Nazarewicz and M.~P\l{}oszajczak,
{\em Phys. Rev. Lett.} (2023) in print
[arXiv:2306.05192 [nucl-th]].

\bibitem{Kegel:2021jrh}
  S.~Kegel, P.~Achenbach, S.~Bacca, N.~Barnea, J.~Bericic, D.~Bosnar, L.~Correa, M.~O.~Distler, A.~Esser and H.~Fonvieille,
  \textit{et al.}
Phys. Rev. Lett. \textbf{130} (2023) no.15, 152502
[arXiv:2112.10582 [nucl-ex]].


\bibitem{Frosch:1968sns}
R.~F.~Frosch, R.~E.~Rand, H.~Crannell, J.~S.~McCarthy, L.~R.~Suelzle and M.~R.~Yearian,
Nucl. Phys. A \textbf{110} (1968), 657-673

\bibitem{Walcher:1970vkv}
T.~Walcher,
Phys. Lett. B \textbf{31} (1970), 442-444


\bibitem{Kobschall:1983na}
G.~Kobschall, C.~Ottermann, K.~Maurer, K.~Rohrich, C.~Schmitt and V.~H.~Walther,
Nucl. Phys. A \textbf{405} (1983), 648-652


\bibitem{Lu:2018bat}
B.~N.~Lu, N.~Li, S.~Elhatisari, D.~Lee, E.~Epelbaum and U.-G.~Mei\ss{}ner,
Phys. Lett. B \textbf{797} (2019), 134863
[arXiv:1812.10928 [nucl-th]].

\bibitem{Lu:2019nbg}
B.~N.~Lu, N.~Li, S.~Elhatisari, D.~Lee, J.~E.~Drut, T.~A.~L\"ahde, E.~Epelbaum and U.-G.~Mei\ss{}ner,
Phys. Rev. Lett. \textbf{125} (2020) no.19, 192502
[arXiv:1912.05105 [nucl-th]].

\end{thebibliography}
\end{document}